\documentclass[sigconf]{acmart}
\AtBeginDocument{%
  }

\setcopyright{none}
\acmDOI{}
\acmISBN{}
\acmConference[LIMITS '25]{11th Workshop on Computing within Limits}{June 26--27, 2025}{Online}
\acmISBN{}

\begin{document}

\title{Civil Servants as Builders: Enabling Non-IT Staff to Develop Secure Python and R Tools}

\author{Prashant Sharma}
\authornote{The author is a Research Data Specialist at the California Department of Food and Agriculture (CDFA) at the time of writing. This paper represents independent research conducted outside the scope of official duties and does not reflect the views, policies, or endorsements of the State of California or the CDFA. The author’s prior experience includes work at a startup that collaborated with government agencies in Canada and the United Arab Emirates. All scenarios and observations presented are hypothetical, based on general knowledge of public sector processes, and do not reflect internal practices of any specific department or agency, including those the author has been affiliated with.}
\email{prashant.sharma@berkeley.edu}
\orcid{1234-5678-9012}
\affiliation{%
  \institution{Independent Researcher}
  \city{Sacramento}
  \state{California}
  \country{USA}
}


\begin{abstract}
Current digital government literature focuses on professional in-house IT teams, specialized digital service teams, vendor-developed systems, or proprietary low-code/no-code tools. Almost no scholarship addresses a growing middle ground: technically skilled civil servants outside formal IT roles who can write real code but lack a sanctioned, secure path to deploy their work. This paper introduces a limits-aware, open-source and replicable platform that enables such public servants to develop, peer-review, and deploy small-scale, domain-specific applications within government networks via a sandboxed, auditable workflow. By combining Jupyter Notebooks, preapproved open-source libraries, and lightweight governance, the platform works within institutional constraints such as procurement rules and IT security policies while avoiding vendor lock-in. Unlike low/no-code approaches, it preserves and enhances civil servants’ programming skills, keeping them technically competitive with their private-sector peers. This contribution fills a critical gap, offering a replicable model for public-sector skill retention, resilience, and bottom-up digital transformation.

\end{abstract}


\maketitle

\section{Introduction}

Government agencies operate under important mandates: to safeguard public resources, ensure procedural fairness, and uphold high standards of security. To meet these obligations, digital tools in the public sector are typically subjected to rigorous procurement procedures and strict IT oversight. These institutional safeguards exist for good reason—they are designed to prevent waste, ensure equitable access to contracting opportunities, and reduce risks to sensitive data and infrastructure. However, they also have a cost: civil servants who identify clear opportunities for internal automation or lightweight tooling often face long procurement cycles or compete for limited IT bandwidth already stretched thin.

Procurement departments, for instance, are responsible for ensuring that every expenditure serves the public interest and that projects do not bypass fair competition. Similarly, IT teams act as custodians of security and stability, ensuring that all deployed tools conform to internal standards and do not introduce cybersecurity vulnerabilities. In environments where time, staffing, and attention are chronically limited, both functions tend toward caution. This tendency, while understandable, can unintentionally suppress creativity, especially when civil servants propose novel solutions that fall outside conventional procurement or IT service delivery pipelines.

Today, many civil servants possess  technical skills, such as scripting in Python or R, querying internal databases, or creating dashboards using open source libraries. However, these skills often remain siloed within individuals or teams. When a civil servant builds a script that automates a complex task or provides interactive visualizations, there is often no sanctioned pathway to scale that solution. Formalizing even a simple application can require navigating approvals, documentation, security reviews, and infrastructure constraints that are disproportionate to the complexity of the tool. As a result, practical innovations built by domain experts often remain trapped in local files or are discarded when the staff member leaves. In addition, some government IT departments may interpret security controls, such as those outlined in NIST SP 800-53 ~\cite{NIST80053r5}, as requiring strict oversight of programming environments, resulting in outright bans on tools like Jupyter Notebook, Python, or R for non-IT teams.

This paper proposes a speculative but feasible system that enables civil servants to safely build and deploy internal tools within institutional limits. In doing so, it gestures toward a broader shift in government digital transformation, moving beyond reliance on external vendors or centralized digital teams. The proposed platform allows civil servants to create applications in Jupyter Notebooks and deploy them as secure, interactive web applications accessible exclusively to authenticated users on the government’s internal network, whether on-site or connecting remotely through a secure VPN gateway. Key features include a curated list of preapproved Python, R, or other relevant language libraries, automated sandbox execution, lightweight peer review, and internal-only web deployment. These mechanisms are intended not to replace institutional oversight, but to make it more scalable, allowing frontline staff to contribute low-risk digital tools while maintaining accountability and information security.

Rather than presenting digital transformation as a process that must be outsourced to vendors or specialized IT teams, this paper argues for a shift in perspective: one in which civil servants already inside the system are empowered to create sustainable, domain-specific applications under constrained conditions. This shift reflects the broader values of the Computing within Limits community, working within institutional, infrastructural, and ecological boundaries while prioritizing resilience, adaptability, and the reuse of local knowledge. The sections that follow articulate the structural barriers to such an approach and how the proposed system works with, rather than around, those constraints.

\subsection{Structural Barriers to Digital Innovation}

Despite growing technical fluency among civil servants, systemic and organizational barriers often prevent frontline innovation from taking root. These barriers are not primarily technological; they are deeply embedded in the structures of public administration, which tend to prioritize procedural integrity, risk minimization, and stability over agility and experimentation.

Digital transformation in the public sector frequently proceeds through external procurement. While necessary for ensuring fairness and fiscal accountability, procurement processes are poorly suited to iterative or exploratory work. Edler et al.~\cite{edler2015procurement} note that procurement regimes emphasize compliance and predictability rather than creativity or impact. This orientation often makes it difficult to contract for lightweight, open-ended tools or small-scale internal prototypes. Proposals that do not fit neatly into formal scopes of work or budget categories may be filtered out early, regardless of their practical utility.

At the same time, government IT departments face pressures to maintain operational stability across critical infrastructure. Bozeman and Kingsley~\cite{bozeman1998risk} describe how public sector organizations develop strong cultures of risk aversion—where the political and reputational consequences of software failure far outweigh the rewards of incremental success. As a result, even technically feasible proposals involving new platforms, packages, or data access methods are often denied. This caution systematically excludes open-source or experimental tools, even when they are popular, well-documented, and widely used in academia or industry.

Furthermore, digital innovation is not analogous to constructing a physical asset, such as a bridge or a building, that can be completed and then handed over for maintenance. Mergel~\cite{mergel2017digital} emphasizes that digital systems require continual refinement, adaptation, and support. When digital services are delivered entirely through third-party contracts, agencies often lack the institutional memory or technical fluency to modify them in response to shifting policy or operational needs. This dependency leads to brittle systems and long procurement delays for even minor updates—an unsustainable model under conditions of resource constraint. 

Legacy systems further intensify the challenge. As noted by the UK House of Commons Public Accounts Committee~\cite{walker2025govai, publicaccounts2025ai}, much of government infrastructure relies on aging software that is difficult to modernize and expensive to maintain. These systems consume the bulk of IT budgets, leaving little room for experimentation or small-scale innovation. Moreover, their rigidity often requires new tools to conform to outdated standards, thereby limiting what frontline developers can feasibly build.

These constraints are not temporary; they are structural. Yet as the Computing within Limits community has emphasized, working within limits does not mean abandoning innovation—it means redirecting it~\cite{Tomlinson2017LimitsGrowth, Penzenstadler2015Collapse}. Digital infrastructure should be made resilient by embedding it in local practice, enabling small teams to operate and adapt their own tools without continuous reliance on external vendors or high-overhead processes. While much of the Computing within Limits community has focused on ecological and material resource constraints of the future, some scholarship argues that the scenarios anticipated by limits community are not just future concerns but present realities in many domains. Chen ~\cite{chen2016strategy} proposes that limits-aware computing should increasingly engage with real, immediate problems to make a difference today while preparing for potential future societal crisis, a perspective in conversation with  by Tomlinson et al.’s ~\cite{tomlinson2012collapse} work on {\itshape collapse informatics}, the study, design, and development of sociotechnical systems in the abundant present for use in a future of scarcity. Addressing structural barriers to small-scale digital innovation in government embodies many characteristics of a limits context. In fact, solving today's deeply embedded limits problems may offer a better foundation for resilience than designing for hypothetical futures—because the latter are difficult to predict. The system proposed in this paper thus contributes to the growing body of work ~\cite{chen2016strategy, nardi2018computing} that views collapse not as a singular future event, but as an ongoing condition that demands adaptive, embedded responses from within existing structures.

\section{System Goals and Assumptions}
The proposed system is designed with several key assumptions and goals:

\begin{itemize}
\item \textbf{No Vendor Lock-In}
A foundational assumption of the proposed platform is the elimination of vendor lock-in and reliance on proprietary low-code environments, which pose significant risks to government IT sustainability. By exclusively using open-source tools like Jupyter Notebooks and deploying applications within secure internal networks, the platform ensures long-term institutional control, modifiability, and resilience without dependency on commercial vendors. Unlike proprietary low-code platforms, which often suffer from opaque licensing, limited long-term support, and inferior documentation and quality compared to open-source ecosystems, this approach leverages the robust, community-driven standards of open-source software.
    \item \textbf{Empower Domain Experts:} 
Civil servants possess deep, situated knowledge of government workflows, policies, and public needs—insights that external vendors often lack. Many already write analysis scripts in languages like Python or R to meet their day-to-day needs, yet systemic barriers prevent this technical ability from being channeled into broader digital innovation. By enabling civil servants to build and deploy tools themselves, the platform recognizes that domain expertise is a critical asset, not something to be outsourced. Unlike traditional vendor-driven development, which often struggles with costly requirement gathering and miscommunication cycles, empowering internal experts allows for iterative, agile development directly informed by frontline realities. In the spirit of Computing within Limits, this approach values local, embedded knowledge over external consultancy models, fostering systems that are better adapted to the environments in which they must operate.

\item \textbf{Interactive Applications:} 
The platform supports the creation of rich, interactive web applications that allow end users to input parameters, explore data, visualize results, and export outputs in a clean interface,  without requiring them to engage with underlying code written in Jupyter Notebook. This system makes civil servants’ technical work accessible to non-programmer colleagues across departments. By transforming notebooks into interactive applications, the system bridges the gap between code and operational tools, allowing for responsive feedback loops and rapid iteration. In constrained environments where formal IT resources are scarce, such lightweight applications offer a practical path to increasing internal digital capacity without the need for heavy, vendor-led IT infrastructure investments.

\item \textbf{Privacy and Security:} 
All applications created on the platform are securely hosted within the government’s internal network, ensuring that sensitive data remains protected and accessible only to authenticated users. To minimize security risks, developers are restricted to a vetted set of preapproved libraries, with a streamlined workflow for requesting and approving new libraries when necessary. This structure acknowledges the legitimate concerns of IT security teams while still promoting flexibility and innovation. Furthermore, cross-departmental collaboration on maintaining the shared library list reduces redundant overhead and accelerates safe experimentation. For example, multiple departments within a state government might collaborate to maintain a shared list of approved libraries. From a LIMITS perspective, this security model accepts the inevitability of constraints—legal, ethical, and infrastructural—and designs innovation pathways that work within them, rather than ignoring them.

\item \textbf{Lightweight Governance:} 
Rather than imposing heavy approval processes typical of traditional IT deployments, the platform emphasizes lightweight, efficient governance mechanisms. Each application undergoes a quick but meaningful peer review, modeled after common software engineering practices like GitHub pull requests, to catch obvious issues and ensure basic quality standards. Version control ensures that any changes are tracked and recoverable, and audit trails log who uploaded, reviewed, and approved each application, ensuring accountability without bureaucratic friction. This approach aligns with LIMITS values by creating governance structures that are resilient and sustainable under conditions of limited administrative and technical capacity. It demonstrates that good governance need not be synonymous with delay.

\item \textbf{Cultural Transformation:} 
By enabling civil servants to build, deploy, and maintain their own digital tools, the platform challenges the entrenched notion that public sector innovation must always flow through formal IT procurement or external vendors. It reframes civil servants as capable technologists who, given the right platforms and safeguards, can drive meaningful change from within. This reorientation not only increases organizational agility but also fosters a culture of continuous learning, creativity, and ownership among public employees. The system also empowers innovative civil servants who do not want to be siloed into learning only low-code proprietary tools, and who may instead wish to gain or refine hands-on coding skills, ensuring they remain as technically competitive as their private sector counterparts. Consistent with LIMITS thinking, the platform recognizes that systemic resilience is best achieved not by relying on external interventions but by cultivating internal capacities that can adapt and evolve under constraint. Civil servants are not passive users of technology; they are, and must be, its stewards.
\end{itemize} 
\section{Related Work}
\subsection{The Gap: Everyday Tool Development by Domain Expert Civil Servants}
While extensive literature covers formal in-house development by professional engineers versus outsourcing to vendors, there exists a significant gap in addressing everyday tool development by domain expert civil servants. This represents a fundamentally different category that mainstream conversations have largely overlooked.
The existing discourse typically frames government technology as a binary choice: hire professional software engineers including those at specialized agencies such as USDS or contract with vendors. ~\cite{mergel2017digital, MergelAgile} However, this ignores a critical middle ground—the small, domain-specific tools that technically skilled civil servants could build to solve immediate challenges. These differ fundamentally from enterprise systems in scope and complexity, yet face the same institutional barriers as major software projects.
A real-world example of addressing this middle ground is the UK Ministry of Justice Analytical Platform (AP), documented in official user guidance~\cite{MoJAnalyticalPlatformUserGuidance} and written evidence to parliament~\cite{MoJ2021WrittenEvidence}. The AP provides civil servants with secure, containerized access to modern programming environments such as JupyterLab and RStudio, enabling analytical workflows while complying with stringent data security requirements.

The practitioner perspective on the AP is captured in a 2018 article by Robin Linacre on the official website of the UK Government~\cite{moj_ap_blog_2018}. Reflecting on his early experience as a government analyst in 2006, Linacre recalled: “we worked mainly with spreadsheets, and sometimes with more specialist proprietary tools. Working with bigger, complex datasets was difficult and software licensing and training was very expensive. Government IT often got in the way, and people joked how we were years behind the private sector.” He further noted that while the past decade has seen “an explosion in free and open source analytical tools” enabling more advanced analysis, “historically these tools have not been available to government analysts. The key difficulty is in giving analysts greater freedom, while safeguarding sensitive government data.” As described earlier in this paper, internal government IT functions, while essential for maintaining security and compliance, are often very risk averse. This tendency, as reflected in Linacre’s account, can inadvertently limit access to modern analytical tools to civil servants. 

While platforms such as the AP address some of the same challenges identified in this paper, their publicly available descriptions are limited to official websites, blog posts, and parliamentary evidence rather than peer-reviewed literature. Moreover, their designs are tailored to the specific organizational context and are not presented as a generalized, open-source reference architecture for adoption across jurisdictions.

To my knowledge, no prior work has combined secure, sandboxed execution; curated package governance; lightweight peer review; and reproducible deployment pipelines into a single, replicable platform specifically aimed at technically skilled but non-IT civil servants operating within government networks. My contribution lies in integrating these elements into a coherent, limits-aware model that can be readily adapted by agencies with similar institutional constraints.

\subsection{End-User Development in HCI}
The concept proposed in this paper is closely related to research on End-User Development (EUD), which focuses on enabling people who are not professional developers to create or adapt software for their own purposes~\cite{lieberman2006enduser}. EUD systems often lower barriers through visual environments, templates, or domain-specific languages, while supporting iterative adaptation by users.~\cite{fischer2004metadesign} The proposed platform follows this tradition by transforming Jupyter Notebooks, already familiar to many civil servants, into secure, interactive applications. While it requires some programming skills in languages like Python, it significantly enhances accessibility for civil servants compared to traditional software development, empowering them to build tailored solutions without the need for extensive technical expertise or complex development processes. 
Recognizing the unique domain expertise of civil servants, prior research in End User Development have proposed platforms to enable public-sector staff to participate in creating e-government services. For example,  Fogli and Provenza ~\cite{fogliEndUserEGov} present a framework that empowers administrative personnel to create e-government services instead of software professionals. However, these systems are primarily oriented toward participants without programming skills, focusing on visual modeling, wizards, and guided workflows. In contrast, the platform proposed in this paper targets a different audience, civil servants who already possess programming skills, and seeks to remove structural and procedural barriers that prevent them from securely deploying the small-scale tools they can already create. 

\subsection{Technical Inspiration}
Technically, the platform draws inspiration from existing open-source tools designed to turn computational notebooks into interactive applications, such as Voilà ~\cite{voila2025} and Mercury~\cite{mljar_mercury_2025}. These tools demonstrate how end-user code artifacts can be repurposed into usable interfaces without requiring professional web development skills. While these systems have been applied in scientific computing, education, and industry, they are rarely adapted to environments with strict security, procurement, and governance constraints. By incorporating controlled library whitelists, sandbox execution, and internal-only deployment, the proposed platform recontextualizes these technical patterns for public-sector use. This adaptation illustrates how established open-source components can be recombined to meet institutional requirements without resorting to proprietary low-code solutions that risk vendor lock-in.

\section{System Overview}
The core philosophy of the system is to work with, rather than against, the institutional limits present in government settings: recognizing that procurement rules, IT restrictions, and limited technical support are enduring realities, not temporary obstacles. The platform’s use of Jupyter Notebooks reflects this ethos: they are already widely used for internal data work, are open-source and auditable, and avoid the risk of vendor lock-in—a critical concern for sustainability in public institutions. Rather than relying on proprietary low-code solutions, the platform prioritizes transparency, modifiability, and long-term institutional control. Its architecture distributes responsibility through peer review and versioning, aligning with open-source development norms and reducing burden on central IT. The system can be built using open-source technologies, such as Docker and Kubernetes, which can be shared as a reusable template across government agencies. \smallskip

\begin{figure}[h]
  \centering
  \includegraphics[width=\linewidth]{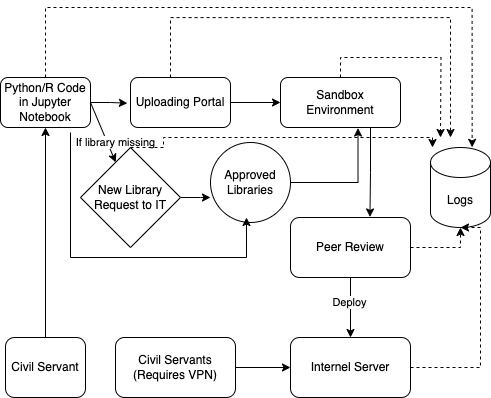}
  \caption{System Architecture for Notebook to WebApp}
  \Description{Visual illustration of the system design, showing the flow from civil servant authorship to final deployment within a secured environment.}
  \label{fig:system_architecture}
\end{figure}

Figure~\ref{fig:system_architecture} illustrates the full pipeline from notebook authoring to live application deployment.

\subsection{Authoring the Notebook}
Civil servants initiate the process by writing Jupyter Notebooks that encapsulate their data analysis, internal API interactions, visualizations, and workflows. The focus remains on domain logic and usability rather than web development. To make notebooks interactive, authors can add lightweight configuration metadata, such as a YAML header specifying which variables should be exposed as user inputs (e.g., dropdown menus, sliders, or file upload fields).

This approach builds on lessons from open-source frameworks like Mercury~\cite{mljar_mercury_2025} and Voilà~\cite{voila2025}, which demonstrate that notebooks can serve as viable, low-code front-end interfaces. 

\subsection{Upload to Portal}
Once the notebook is complete, the civil servant uploads it to the platform’s secure internal portal. This portal acts as the primary interface for managing application submissions, status updates, version history, and peer reviews.

Upon upload, the platform performs an automated validation process:

\begin{itemize}
\item \texttt{Checks Package Whitelist}: Authors must submit a dependency file (e.g., \texttt{requirements.txt} for Python) listing all external libraries their notebook depends on. The platform cross-checks these against a curated registry of preapproved libraries. If any unapproved packages are detected, the author is prompted to submit a streamlined request for IT security review.
\item \texttt{Commits Version Control}: Each notebook upload or revision is committed to an internal Git-based version control system. This ensures that all changes are tracked, historical versions are preserved, and any previous version can be restored if necessary.
\end{itemize}

By frontloading dependency checks and automating version management, the platform reduces the administrative burden on IT departments while maintaining security and traceability standards.

\subsection{Automated Sandbox Build and Test}
After successful upload, the platform automatically spins up a secure sandbox environment to validate the notebook execution:

\begin{itemize}
\item \texttt{Execution and Rendering}: The notebook runs end-to-end in a containerized environment to generate outputs such as plots, tables, and downloadable data files. Raw code cells are hidden from the final interface, ensuring users interact only with the intended front-end elements.
\item \texttt{UI Generation}: Based on notebook metadata or YAML configurations, the system dynamically generates user interface controls. For example, a variable annotated as a dropdown choice will automatically appear as a selection menu in the resulting app. This feature transforms notebooks from static documents into parameter-driven, interactive applications.
\item \texttt{Sandbox Restrictions}: The execution environment strictly limits external internet access and enforces read-only permissions for database credentials. Only connections to vetted internal APIs are permitted. These security measures minimize the risk of data leakage or unauthorized external communication, adhering to public sector IT security best practices.
\end{itemize}

If any failures occur, such as missing dependencies, runtime errors, or unauthorized operations, the platform halts the process and provides detailed feedback to the author for revision.

\subsection{Peer Review and Approval}
Following a successful sandbox build, the system initiates a lightweight peer review process:

\begin{itemize}
\item A designated peer reviewer, typically a colleague within the same program area, is notified.
\item Reviewers are asked to validate basic functionality (e.g., does the app run correctly and produce expected outputs?) rather than performing deep technical security audits.
\item Review actions—such as approval, request for changes, or rejection—are logged in the platform, creating an auditable trail for internal governance and compliance purposes.
\end{itemize}

This peer review model draws inspiration from collaborative software engineering practices, balancing quality assurance with efficiency. It supports rapid deployment while ensuring that at least two people assess each tool before it becomes widely accessible.

\subsection{Deployment to Internal Server}
Upon approval, the notebook is automatically deployed as a standalone web application on the government’s internal servers. Key deployment features include:

\begin{itemize}
\item Applications are assigned stable internal URLs (e.g., \url{https://apps.department.gov/internal/<app-name>}), easily shareable within teams.
\item Access is restricted to authenticated users connected through the intranet or VPN.
\item Each application runs inside its own isolated container, ensuring that resource consumption, crashes, or security vulnerabilities in one app do not affect others.
\item Applications can be scaled horizontally if usage grows, allowing for basic load balancing without requiring civil servants to manage infrastructure directly.
\end{itemize}

Because the deployed applications are fundamentally Jupyter-based backends rendered through secure web frontends, they can evolve easily over time. Civil servants can update their notebooks, submit new versions through the portal, and roll out iterative improvements without requiring full redeployments or external vendor involvement.

This deployment strategy minimizes the operational burden on central IT while enabling frontline teams to maintain, adapt, and enhance their digital tools autonomously—a critical capability for sustainability within the practical limits of government environments.

\section{User Scenario Example}
To illustrate how the platform might function, consider the following hypothetical cases:

\subsection{Spreadsheets Generator – Binita’s Tool}

Binita, a Transportation Engineer at a government department, develops a Jupyter Notebook to automate the generation of Excel spreadsheets. Her tool extracts data from an internal SQL Server and an ArcGIS geodatabase, processes it, and outputs multiple structured Excel files. Previously, preparing these spreadsheets required approximately eight hours of manual work every week; her script reduces the task to just a few seconds.

With the proposed platform, Binita uploads her notebook.
She specifies two configurable inputs—{\itshape month} and {\itshape county name}—using a YAML header. Her notebook relies mostly on preapproved libraries ({\itshape pandas}, {\itshape numpy}, {\itshape geopandas}) but also requires {\itshape spacy}, which is not yet on the approved list. The platform detects the use of an unapproved package and notifies Binita. Binita submits a package approval request for {\itshape spacy} through the platform. The IT security team reviews and approves the {\itshape spacy} library.
The platform sandbox executes her notebook, generates the expected outputs, and configures UI widgets for the specified parameters.
Binita is notified that the tool has passed automated checks and is ready for peer review.

During peer review, Yaw accesses the deployed app through a private preview link.
Yaw suggests clarifying the title of one of the output charts for better interpretability.
Binita updates her notebook accordingly, resubmits the new version, and passes the second review.
The platform redeploys the finalized app and assigns it an internal URL.

Binita shares the link with her team. Team members, including those without coding skills, can now instantly generate customized spreadsheets for different months and counties, significantly improving efficiency.

\subsection{Text Analysis Tool – Sirak’s Tool}

Sirak, a Program Data Specialist at the same government department as Binita but in a different team, develops a Jupyter Notebook that processes textual data and generates outputs based on user input. His tool utilizes the {\itshape pandas} and {\itshape spacy} libraries, both of which are already included in the agency’s preapproved list.  {\itshape spacy} was added after Binita's request while she was building Spreadsheet Generator tool, so it is already available to Sirak.

With the proposed platform, Sirak uploads his notebook. He specifies one parameter, {\itshape day}, as a configurable input using a YAML header. His notebook uses only preapproved libraries ({\itshape pandas}, {\itshape spacy}), so it passes the package validation automatically. The platform sandbox runs his notebook, generates an interface, and sets up a basic text input field for the parameter. Sirak is notified that the tool has passed initial checks and is ready for peer review.

During peer review, Marina reviews the deployed app via a private preview link. Marina suggests that instead of a free-text input, a dropdown list with days of the week (Monday, Tuesday, etc.) would standardize user input. She also recommends clarifying the input label to "Day of Week" to avoid confusion with specific calendar dates. Sirak updates his notebook’s YAML configuration to replace the text input with a dropdown list and clarifies the label. He reuploads the revised notebook, and the platform sandbox re-executes it successfully. The app passes the second peer review without further issues. The platform deploys the updated app with an internal URL.

Sirak shares the link with his colleagues. Staff across the department can now easily select the day of the week from a dropdown menu and run the analysis consistently, without risk of input errors or needing any technical assistance.\bigskip

The entire process, from code upload to internal deployment, can be completed in one to three days, rather than requiring months of procurement, contracting, or IT development time. These scenarios demonstrate how civil servants can deliver rapid, secure innovation within existing structural limits, freeing IT departments from routine application requests while maintaining public sector standards of security, accountability, and fairness.

\section{Limitations and Future Work}
\subsection{Setup Complexity}
Setting up such a system may require formal approvals, IT coordination, and, in some cases, vendor procurement. This raises a valid concern: does the platform merely shift complexity rather than reduce it? To some extent, yes—the initial setup is not trivial. However, unlike bespoke tools or one-off vendor systems, the platform is designed as a reusable template. Once piloted in a few government contexts, it gains institutional legitimacy, enabling other agencies to reference prior deployments and streamline their own approval processes. Early adopters thus pave the way for risk-averse institutions to adopt the platform with greater confidence. Future work could explore shared deployment kits, inter-agency collaborations, or public documentation to formalize this pattern.

\subsection{Small Scale Tools}
The platform is not intended for large-scale software development. It focuses on small, domain-specific tools, such as data cleaning scripts, report generators, or internal dashboards, that many civil servants already create informally or are capable of building but are restricted by IT security policies. By lowering barriers for these lightweight applications, the platform addresses a specific need but does not replace the  need for enterprise-grade systems or formally procured software contracts.

\subsection{Accessibility for Non-Programmers}
In its current form, the platform may exclude non-programmer innovators. Many frontline public servants have valuable ideas for improving workflows but lack coding experience. While Jupyter Notebooks reduce the barrier compared to traditional web development, they still require Python or R fluency. However, this mirrors the status quo rather than creating new obstacles. The platform may even widen participation by fostering collaboration between domain experts and technically skilled colleagues. Even if only one team member knows how to script, they can build tools that automate manual tasks for the entire team, making everyone’s work easier and more efficient.

\subsection{Library Approval Delays}
Although the platform simplifies security governance through sandboxing and package whitelisting, approving new libraries can still cause delays. IT departments, especially in sensitive policy or regulatory environments, are understandably cautious. No system can ensure complete safety, and there remains a risk of introducing vulnerabilities through third-party packages.

\subsection{Feasibility for Small Governments}
While designed for broad use across agencies, the platform may be challenging for smaller local governments with limited IT resources. Maintaining the infrastructure for sandboxed execution, peer review governance, and package registry curation demands sustained organizational capacity. Collaboration across multiple small governments could help pool resources, but further research is needed to assess such models in practice.

\section{Discussion}
Government digital transformation has followed several dominant approaches, each shaped by the institutional and technical constraints of its time. Early systems were often developed in-house by government IT departments as large, centralized projects—highly customized but expensive, inflexible, and difficult to evolve. Over time, many governments shifted toward vendor-driven procurement, outsourcing software development to external contractors. While this model promised efficiency and risk reduction, it often introduced long delays, rigid contracts, and deep dependency on outside vendors for even minor changes.

To address the limitations of these models, many governments introduced centralized digital service teams, such as the UK’s Government Digital Service (GDS), the U.S. Digital Service (USDS), and similar units elsewhere.~\cite{MergelAgile} These teams brought technical talent inside the public sector and demonstrated the value of agile, user-centered design. However, their capacity is often concentrated on selected high impact projects, not the small, domain specific tools needed by front line staff in everyday operations. More recently, low-code and no-code platforms have been promoted as a way to decentralize development, but they often come with risks of vendor lock-in, limited transparency, and unclear long-term support.

By contrast, in the private sector, many of these challenges
are less pronounced. Teams across industries often have the
flexibility to adopt freely available tools, low-cost software-
as-a-service (SaaS) products, or internal scripts without the
same legal, procurement, or security hurdles. For example, a marketing team might install a browser plugin, automate a spreadsheet, or connect a database to a visualization tool in just a few hours. In government, those same actions could require weeks or months of legal review, procurement paperwork, IT security vetting, or approvals from multiple departments. This stark difference highlights the need for systems that are specifically designed to work within public-sector constraints, rather than assuming the ease and flexibility that private-sector teams often take for granted.

The system proposed in this paper offers a complementary model to existing approaches, not a replacement. It does not reject centralized teams or vendor partnerships. This model recognizes a gap that neither vendors nor specialized digital units can bridge: deep, situated knowledge of frontline workflows and operational nuance. While centralized teams may have advanced engineering or data science skills, they often lack the granular, often tacit knowledge possessed by civil servants working directly on the problem. Empowering those workers to build tools does not just increase efficiency, it improves relevance, adaptability, and ownership.

By standardizing a reusable architecture, the platform lowers the barrier for risk-averse agencies to adopt it over time. Early implementations generate institutional precedent, allowing others to follow with less friction.

An important consideration in evaluating the long-term impact of the platform is that it will enable a mix of types of work: some applications will automate entirely new tasks that were previously performed manually or not at all, while others may take over small-scale functions once handled by vendors or centralized IT teams. Success can be measured by tracking productivity improvements in formerly manual tasks, as well as reductions in vendor reliance or IT requests for small-scale tool development. In addition, qualitative research, including interviews and case studies with civil servants using the platform, can provide valuable insights into how these tools enable new ways of working and influence organizational culture.

\subsection{Building Resilience Through LIMITS}
From a LIMITS perspective, this approach reflects the need to build within real constraints. The system uses what institutions already have: public servants with technical skills, open-source tooling, and secure internal networks. Designing for institutional resilience today, within bureaucratic and infrastructural limits, may offer a more grounded foundation for navigating future scenarios in other sectors that presently enjoys more flexibility on using low cost SaaS tools, open source packages, or internal scripts with minimal oversight. As LIMITS scholars have noted~\cite{Tomlinson2017LimitsGrowth, Penzenstadler2015Collapse, chen2016strategy}, future scenarios shaped by potential ecological instability, economic degrowth, supply chain fragility, or geopolitical disruptions could impose new limits similar to those faced by governmental organizations in other sectors. In that context, innovating within these constraints now may offer practical patterns and infrastructure that prove valuable far beyond their original use case.

\section{Conclusion}
This paper proposes a speculative but feasible platform that empowers civil servants to build and deploy internal tools safely within the structural limits of government work. By providing a secure, auditable pipeline from notebook to web application, the system upholds core public sector values of accountability, security, and fairness, while dramatically increasing agility at the frontline.
Rather than bypassing procurement or IT processes, the platform complements and supports them, freeing critical resources for higher-risk projects and enabling routine innovation to flourish. It repositions civil servants not just as users of technology, but as active builders of digital solutions.

\bibliographystyle{ACM-Reference-Format}
\bibliography{sample-base}

\end{document}